# Cooperative motions in a finite size model of liquid silica: an anomalous behavior

Victor TEBOUL

Laboratoire des Propriétés Optiques des Matériaux et Applications, CNRS UMR 6136, Université d'Angers, 2 Bd Lavoisier, F-49045 Angers, France

E-mail: victor.teboul@univ-angers.fr

**Abstract:**

Finite size effects on dynamical heterogeneity are studied in liquid silica with Molecular Dynamics simulations using the BKS potential model. When the system size decreases relaxation times are found to increase in accordance with previous results in finite-size simulations and confined liquids. It has been suggested that this increase may be related to a modification of the cooperative motions in confined liquids. In agreement with this hypothesis we observe a decrease of the dynamical heterogeneities associated to the most and the least mobile atoms when the size L decreases. However we find that the decrease of the dynamical aggregation associated to the least mobile atoms is much more important than the decrease associated to the most mobile atoms. This result is surprising as the liquid is slowed down. The decrease of the heterogeneous behavior is also in contradiction with the increase of the heterogeneities observed in liquids confined in nanopores. However an increase of the non-Gaussian parameter appears both in nanopores and in the finite size simulations. As the non-Gaussian parameter is usually associated with dynamical heterogeneities, the increase of the non-Gaussian parameter together with a decrease of dynamical heterogeneity is also surprising.

**Introduction:**

The existence of cooperative molecular motion in supercooled glass-forming liquids is commonly invoked [1] as the likely explanation for the dramatic increase of the viscosity as the liquid is cooled toward its glass transition. Cooperative motions associated with an heterogeneous dynamics are also commonly postulated in order to explain the non-exponential behavior of correlation functions and the non-Arrhenius behavior [2] with temperature of the viscosity of most glass forming liquids. Dynamical heterogeneities and associated cooperative behavior have been reported either experimentally near the glass transition temperature or with molecular dynamics simulations well above this temperature [2,3]. From MD simulations these heterogeneities are usually characterized by an aggregation of the most mobile atoms [2-11] and of the least mobile atoms [11]. Because the structure changes only slightly when the temperature of the liquid decreases, the increase of the cooperative motions expected by most theories has been associated with the observed dynamical heterogeneities. Whether dynamical heterogeneities are partly at the origin of the strange comportment of glass-forming liquids or are only an interesting consequence of it, is however still a matter of conjecture.

When the liquid is confined inside a pore a few nanometers across the cooperative motions will not be able to extend to distance larger than the pore size. As a consequence, a modification of the dynamical properties is expected which may give information on the nature of the cooperative motions [2,3,9,12,13]. Indeed, the physical properties of the liquid change drastically with confinement [9,12,14,15]. However if these changes originate partly from finite size effects, anisotropy and surface effects also play an important role [12,13]. In other words, liquids confined in nanopores are subject to different contributions: strong surface contribution and finite-size contribution are both present [12,14]. It has been found previously [9,12,14] that surface contribution is high as expected near the surface of the pore and decrease exponentially [14] when the distance from the surface increases. In opposition with finite-size effects that will not depend on the distance from the surface of the pore. As a result when pore diameters are large enough to eliminate surface effects and short enough for non negligible finite size effects, finite-size effects are expected to predominate in the center of the pore. In agreement with this viewpoint, it has been observed for various liquids that the alpha relaxation times may be fitted by the sum of an exponential decreasing from the pore surface (surface effect) and a constant part (finite-size effect) [12].

In this context, a direct investigation of finite-size effects may help to separate the different contributions of the above-mentioned mechanisms that appear when a liquid is confined into a pore. These studies may also give indirect information on the size of cooperative motions in the liquid. And more generally give information on the size of the phenomena of physical importance for the relaxation mechanisms in the liquid. A simple method, which seems not being affected by surface effects is to use a finite cubic box of size L with periodic boundary conditions (p.b.c.). In opposition with confined liquids dynamics that displays an increase or a decrease of relaxation times due to surface contributions [14] depending on the roughness of the surface; for finite boxes with p.b.c simulations, when the size L decreases, it has been found [16-18] in various liquids, including silica [17,18] that relaxation times increase. This increase was found to be larger for strong liquids [18] like silica than for fragile liquids [19]. The reason for this increase is however still unclear. It has been suggested [16] that this increase may be related to a modification of the cooperative motions in confined liquids. In agreement with this hypothesis we observe a decrease of the dynamical heterogeneities associated to the most and the least mobile atoms when the size L decreases. However we find that the decrease of the dynamical aggregation associated to the least mobile atoms is much more important than the decrease associated to the most mobile atoms. This result is surprising as the liquid is slowed down. The decrease of the heterogeneous behavior is also in contradiction with the increase of the heterogeneities observed in liquids confined into nanopores [9]. Meanwhile an increase of the non-Gaussian parameter appears both in nanopore confinement and in the finite size simulations. As the non-Gaussian parameter is usually associated with dynamical heterogeneities, the increase of the non-Gaussian parameter together with a decrease of dynamical heterogeneity is also surprising. Finally, it has been found previously in silica [20] that the Kohlrausch-Williams-Watts (KWW) parameter associated to the intermediate scattering function does not show significant finite-size effects. Because an heterogeneous dynamics increases the stretching, the KWW parameter (or stretching parameter) has been linked to the observed heterogeneous dynamics. In this viewpoint a decrease of the heterogeneity has then to increase this parameter (i.e. to decrease the stretching). That the KWW parameter doesn't show significant finite-size effects seems then to show that another cause compensate the decrease of the heterogeneity for the stretching parameter evolution.

**Calculation:**

Our simulations use the Verlet algorithm [21] to solve the equations of motions with the BKS potential [22,23]. This potential was reported to be one of the most realistic potential in silica for dynamical effects [24]. The time step was chosen equal to $10^{-15}$ s. The reaction field method [21] was employed to take into account long-range electrostatic interactions in the same conditions than described in ref.[8,10,25]. The simulations are aged during 20 nanoseconds in order to insure stabilization before any treatment. In order to study the size effects we have used various box sizes ranging from 20 Å to 50 Å wide. The density was set constant in our simulations at 2.3 g/cm3.

In the Markovian approximation, the self-part of the Van Hove correlation function $G_s(r,t)$ has a Gaussian form. This function is defined by

$$G_s(\mathbf{r},t) = <\frac{1}{N}\sum_{i=1}^{N}\delta(\mathbf{r} + \mathbf{r}_i(t_0) - \mathbf{r}_i(t + t_0))> \qquad (1)$$

and $4\pi r^2 G_s(r,t)$ represents the probability for a particle to be at time $t+t_0$ at a distance $r$ from its position at time $t_0$. Departure from this Gaussian form has been found in various glass forming liquids and is thought to be due to dynamical heterogeneities. Such deviations are usually characterized by the Non-Gaussian parameter:

$$\alpha_2(t) = 3 <r^4(t)> / 5 <r^2(t)>^2 \; - 1 \qquad (2)$$

where $<r^2(t)>$ is the mean square displacement.

We define the mobility $\mu_{i,t0}(t)$ of atom i at time $t_0$ within a characteristic time t, by the relation:

$$\mu_{i,t0}(t) = |\mathbf{r}_i(t + t_0) - \mathbf{r}_i(t_0)| / (<r^2(t)>)^{0.5} \qquad (3)$$

The mobility of atom i at time $t_0$ is then defined as the normalized displacement of atom i during a time t. We will omit in further discussion the time $t_0$ which will disappear in the mean statistical values. We then select atoms of high or low mobility for the calculation of dynamical heterogeneity. This selection is then dependent on the time t chosen for the definition of the mobility $\mu_i(t)$. We define here as most mobile (MM) the 6 percent atoms with largest mobility, and as least mobile (LM) the 6 percent atoms with lowest mobility. We then select atoms of high and low mobility for the calculation of dynamical heterogeneity. This selection of atoms of high and low mobility depends on the time t chosen in the definition of the mobility. We define here the function:

$$A^+(r,t) = G_{mm}(r,0)/G(r,0) - 1 \qquad (4)$$

In this formula $G_{mm}(r,0)$ is the radial distribution function between the most mobile oxygen atoms, and $G(r,0)$ is the mean radial distribution function between two oxygen atoms. $A(r,t)$ gives a measure of the correlation increase between mobile atoms. Similarly we define $A^-(r,t)$ for the least mobile atoms. In order to eliminate infinite values, we define $A(r,t)=0$ when $G(r,0)<0.05$. This definition implies that $A(r,t)$ is only meaningful above a certain cut-off distance defined by $G(r,0)>0.05$. This cut-off distance is here 2.2 Å for oxygen atoms.

We then define the integrals $I^{+/-}(t)$ of the functions $A^{+/-}(r,t)$ by:

$$I^{+/-}(t) = \int_0^{R_C} A^{+/-}(r,t).4\pi r^2 dr \qquad (5)$$

In our simulations $A(r,t)$ is only defined for $r < L/2$, and the box size L evolves from 20 to 50 Å. In order to eliminate direct truncation effects on the evolution of the function I(t) with the size of the box, we have then truncated the integral in the following calculations at a cut-off value $R_C$ =10 Å which corresponds to the shorter L/2 value investigated. In our notations functions $A^-(r,t)$ and $I^-(r,t)$ correspond to the least mobile atoms while functions $A^+(r,t)$ and $I^+(t)$ correspond to the most mobile atoms. Functions A(r,t) represent the correlation increase between atoms of approximately the same mobility, and distant of *r*. Functions I(t) represent then the global increase of the correlation between atoms of high ($I^+$) or low ($I^-$) mobility. Following ref.[11] we will name the function I(t): Intensity of the aggregation.

**Results:**

Figure 1 shows the non-Gaussian parameter $\alpha_2(t)$ (2) and the mean square displacement $<r^2(t)>$ of oxygen atoms, for various system sizes L ranging from 20 Å to 50 Å. We observe in figure 1 an increase of the non-Gaussian parameter when the system size L decreases. As usual in glass-forming liquids, this increase of the non-Gaussian parameter is observed together with a slowing down of the dynamics of the liquid [12]. This slowing down has been reported previously by different authors [17,18]. In particular it has been demonstrated [18] that the short time dynamics (for t < 0.1 ps ) was not affected by the system size modification while the long time dynamics slows down. In order to illustrate this evolution, in the inset of Figure 1 we have plotted the mean square displacement evolution with the system size at a temperature of 3100 K. We observe in the inset that the modification of the mean square displacement appears around the end of the plateau time regime. This result suggests that the system size modification affects the probability for an atom to escape the cage constituted by its neighbors [26]. Because dynamical heterogeneity are also maximum around the end of the plateau time regime, this result agrees well with an interpretation of size effects resulting from a modification of cooperative motions [16]. This result may be seen clearly in figure 1 which displays the non Gaussian parameter time evolution together with the mean square displacement time evolution, because the non Gaussian parameter is usually considered as a measure of the dynamical heterogeneity. In Figure 1 we observe indeed that the maximum of the non-Gaussian parameter corresponds to the time regime where the mean square displacements modification begins. On the other hand the observed increase of the non-Gaussian parameter seems to be in contradiction with the expected decrease of the cooperativity when the system size decreases.

We also observe in Figure 1 that the non-Gaussian parameter increases strongly as the half system size (L/2) decreases below 12.5 Å. In parallel we observe that the mean square displacements changes also strongly when the same system size is crossed. These results suggest that the size of cooperative motion at the temperature of study (3100 K) is around 12.5 Å. The characteristic time $t^*$ which corresponds to the maximum of the non-Gaussian parameter also increases for the same system size, but more slightly as seen in figure 1. We note that this increase of $t^*$ is much weaker than what will be observed with a decrease in temperature corresponding to the same maximum value of the non-Gaussian parameter. A slowing down of the dynamics associated with an increase of the non-Gaussian parameter may also be observed when supercooled liquids are confined inside pores a few nanometers across [12]. However the increase of the non-Gaussian parameter and the dynamical slowing down are much larger in nanopores confinement than what is observed here for the same sizes of confinement. In agreement with previous studies of various surfaces contributions [14], this result suggests that the slowing down of the dynamics and the associated increase of the non-Gaussian parameter is largely due in nanopores to the interaction with the surface of the pore.

In supercooled liquids the non-Gaussian parameter (2) usually increases when the temperature decreases and this evolution is associated with an increase of the dynamical heterogeneity. The time evolution of the non-Gaussian parameter corresponds also to the time evolution of dynamical heterogeneity. These comportments have been observed in particular in liquid silica [8,10] and in various glass-formers. For these reasons among others, in glass-forming liquids the non-Gaussian parameter is usually associated with the presence of dynamical heterogeneity. We will now investigate the evolution of the dynamical heterogeneity when the system size decreases. When the system size decreases, the dynamical aggregation will not be able to extent to distances larger than the system size, and this truncation effect leads to a decrease of the dynamical heterogeneity. In other words a decrease of $R_c$ due to the decrease of the box size, in equation (5) leads to a decrease of I, because A is positive. Then the question of another non-trivial effect on the heterogeneity arises. We will then focus our attention on this question (i.e. on the evolution of the heterogeneity with a fixed cut-off $R_c$).

Figures 2a and b show the evolution of the dynamical heterogeneity when the system size decreases. In this calculation in order to eliminate the simple truncation effect mentioned above, on the size of the heterogeneities, we have used the same cut-off value $R_c=10$ Å in the calculation of the intensities $I^+(t)$ or $I^-(t)$ (5) for the different system sizes. This value of $R_c$ corresponds to the half-size of the shorter system considered here. We observe that even without the truncation effect, the dynamical heterogeneities decrease when the system size decreases. Surprisingly, this decrease is here associated to an increase of the non-Gaussian parameter. This result is amazing if we suppose that the non-Gaussian evolution of the van hove correlation function (1) that results in a non-Gaussian parameter different from zero, is entirely due to the heterogeneity in displacements. A decrease of the heterogeneous motion will then lead to a decrease of the non-Gaussian parameter in opposition with what is observed here. This result suggests that the non-Gaussian behavior is not only due to the presence of the dynamical heterogeneity but that other causes are also involved. We note that in the case of a liquid confined in a pore an opposite result is observed [9]. In this case the heterogeneity increases together with the non-Gaussian parameter when the liquid is confined. This increase of the heterogeneous motion is however maximum around the pore surface and may then be related to surface effects.

In Figure 2a we show for comparison the non-Gaussian parameter corresponding to the largest system size investigated (L=50Å) i.e. to the liquid dynamics without finite-size effects. The Figure shows that the Non Gaussian parameter follows the main time evolution of the intensity $I^+(t)$ of the heterogeneity (aggregation of the most mobile atoms). This result suggests a relation between the non-Gaussian parameter and the dynamical heterogeneity as observed in supercooled water previously [11]. We also observe in Figure 2a different maxima in the intensity of the heterogeneity $I^+(t)$. These maxima do not appear for the intensity $I^-(t)$ associated to the aggregation of atoms of low mobility in Figure 2b. The first peak corresponds to short times (around 0.4 picosecond). This time regime corresponds to the very beginning of the plateau time regime observed in the mean square displacement. We have to notice that like others authors [27] we also observe a peak in the Non Gaussian parameter but for slightly shorter time scales which may be related to the Si-O bond breaking process. The second peak in Figure 2a corresponds approximately to the maximum of the non-Gaussian parameter. It corresponds to the maximum of the short-range correlation for the dynamical aggregation (this result is observed on the $A^+(r,t)$ functions first maximum which corresponds to the first neighbor position is maximum for $t=t^*$). The third peak corresponds to an increase of the number of atoms correlated at longer distances. This third peak is deeply affected by the cut-off distance $R_c$. When $R_c$ decreases this peak decreases sharply. This result suggests that long range correlations are partly at the origin of this peak. This peak is not observed in supercooled water [11] a liquid which presents however a number of similarities with silica.

We observe in Figure 2a that size effects on the aggregation of the most mobile atoms is weak, for system sizes above or equal to 12 Å, when the direct truncation effect are removed as it is the case in Figure 2a and 2b. The size effects begins then to be important for sizes shorter or around 10Å. In opposition size effects appear around 20 Å for the dynamical aggregation associated to atoms of low mobility as seen in Figure 2b. The heterogeneity associated to atoms of low mobility appear then to be more sensitive to finite size effects, independently of the direct truncation effect associated to the size of the heterogeneity.

A comparison between figure 2a and 2b shows that the aggregation of the least mobile atoms decreases much faster with the system size than the aggregation of the most mobile atoms. This result is surprising as it is associated with a slowing down of the dynamics. If the modification of the dynamics is due to a modification of the cooperative motions associated with dynamical heterogeneity then we expect that a slowing down of the dynamics will be associated to a decrease of the cooperative motion of the most mobile atoms or an increase of the cooperative motion associated to the atoms of low mobility. This result suggests then that the modification of the dynamics is only partly due here to the dynamical heterogeneity or alternatively that the atoms of low mobility have a much weaker effect on the dynamics than the atoms of high mobility.

In order to answer this question we will now investigate the evolution of the self Van Hove correlation function (1) for different system sizes. Figure 3 shows the Van Hove $G_s(r,t^*)$ multiplied by a factor $4\pi r^2$ and corresponding to the system sizes of L/2 = 10 Å and L/2 = 25 Å. We observe in Figure 3 the tail of the function which corresponds to atoms of large mobility, and has been usually associated with dynamical heterogeneity. Due to the presence of this tail the Van Hove correlation function cannot be a Gaussian and the non-Gaussian parameter is then different from zero.

We observe in Figure 3 the self Van Hove correlation functions corresponding to two different box sizes at two different times. The shorter box results are displayed with dashed lines and the larger box results with continuous lines. A short dashed line has been added in order to visualize the first neighbor distance between oxygen atoms ($r=r_{first\ neighbor}$). We observe that two kind of motions are present: The usual continuous diffusive motion which is characterized by a widening of the Van Hove with time and a continuous shift of the maximum of $4\pi r^2 G_s(r,t)$ to larger $r$ values; and the hopping motions which are characterized by an increase of the Van Hove for $r=r_{first\ neighbor}$ leading to the apparition of a bump at this distance and a simultaneous decrease of the first peak. We observe in Figure 3 that the continuous lines curves for r< 2 Å display larger shifts to large r values showing that the continuous flow motion is more important for large boxes than for short boxes. We also observe that the bump for $r=r_{first\ neighbor}$ is roughly the same for the two different boxes.

This result suggests that the hopping motions are less affected by the system size than the sluggish heterogeneous dynamics which is in turn deeply affected. This result may explain the association of an increase of the non-Gaussian parameter with a decrease of the dynamical aggregations. The hopping motions appear indeed as cooperative non-Gaussian motions but with a much smaller aggregation of the mobile atoms than the heterogeneous dynamics.

**Conclusion:**

We have studied finite size effects on dynamical heterogeneity in liquid silica with Molecular Dynamics simulations using the BKS potential model. When the system size decreases relaxation times are found to increase in accordance with previous results in finite-size simulations and confined liquids. It has been suggested that this increase may be related to a modification of the cooperative motions in confined liquids. In agreement with this hypothesis we observe a decrease of the dynamical heterogeneities associated to the most and the least mobile atoms when the size L decreases. However we find that the decrease of the dynamical aggregation associated to the least mobile atoms is much more important than the decrease associated to the most mobile atoms. This result is amazing as the liquid is slowed down. The decrease of the heterogeneous behavior is also in contradiction with the increase of the heterogeneities observed in liquids confined in some nanopores. However an increase of the non-Gaussian parameter appears both in nanopores and in the finite size simulations. As the non-Gaussian parameter is usually associated with dynamical heterogeneities, the increase of the non-Gaussian parameter together with a decrease of dynamical heterogeneity is also surprising. This result seems to eliminate dynamical heterogeneity as the unique cause for the non-Gaussian behavior of glass-forming liquids. Care has then to be taken when using the non-Gaussian parameter as a measure of the heterogeneity in glass-forming liquids.

**Figure caption:**

**Figure 1 inset:** Mean square displacement for oxygen atoms (Å$^2$) in liquid silica for different system sizes. The temperature is 3100K. From top to bottom: Continuous line: L=50 Å (N=9000 atoms), short dashed line (superimposed on the continuous line) L=40 Å (N=4608 atoms), bold dashed line: L=25 Å (N=1125 atoms), dotted line: L=20 Å (N=576 atoms). <r$^2$(t)> is plotted in logarithmic scale.

**Figure 1:** Non Gaussian parameter for oxygen atoms in liquid silica for different system sizes. The temperature is 3100K. From bottom to top: Continuous line: L=50 Å (N=9000 atoms), short dashed line L=40 Å (N=4608 atoms), bold dashed line: L=25 Å (N=1125 atoms), dotted line: L=20 Å (N=576 atoms).

**Figure 2a:** Function $I^+(t) = \int_0^{R_C} A^+(r,t).4\pi r^2 dr$ with R$_c$=10 Å, for the oxygen atoms. Different system sizes are considered: Full circles: L=50 Å (N=9000 atoms), continuous line L=40 Å (N=4608 atoms), empty circles: L=25 Å (N=1125 atoms), triangles: L=20 Å (N=576 atoms). The non-Gaussian parameter (dotted line) multiplied by a factor 1000 and corresponding to the largest size investigated (L=50 Å) is plotted for comparison. The temperature is 3100K.

**Figure 2b:** Function $I^-(t) = \int_0^{R_C} A^-(r,t).4\pi r^2 dr$ with R$_c$=10 Å, for the oxygen atoms. Different system sizes are considered: Full circles: L=50 Å (N=9000 atoms), continuous line L=40 Å (N=4608 atoms), empty circles: L=25 Å (N=1125 atoms), triangles: L=20 Å (N=576 atoms). The temperature is 3100K.

**Figure 3:** Van Hove correlation function $G_s(r,t)$ versus distance r, for the largest and smallest size investigated. Dashed lines: L = 20 Å (N=576 atoms) with t = 0.2 ns (left hand side) and t = 1.75 ns (right hand side); Continuous lines: L = 50 Å (N=9000 atoms) with t = 0.1 ns (left hand side) and t = 1 ns (right hand side). The small dashed line shows the position of the first oxygen neighbor.

Figure 1:

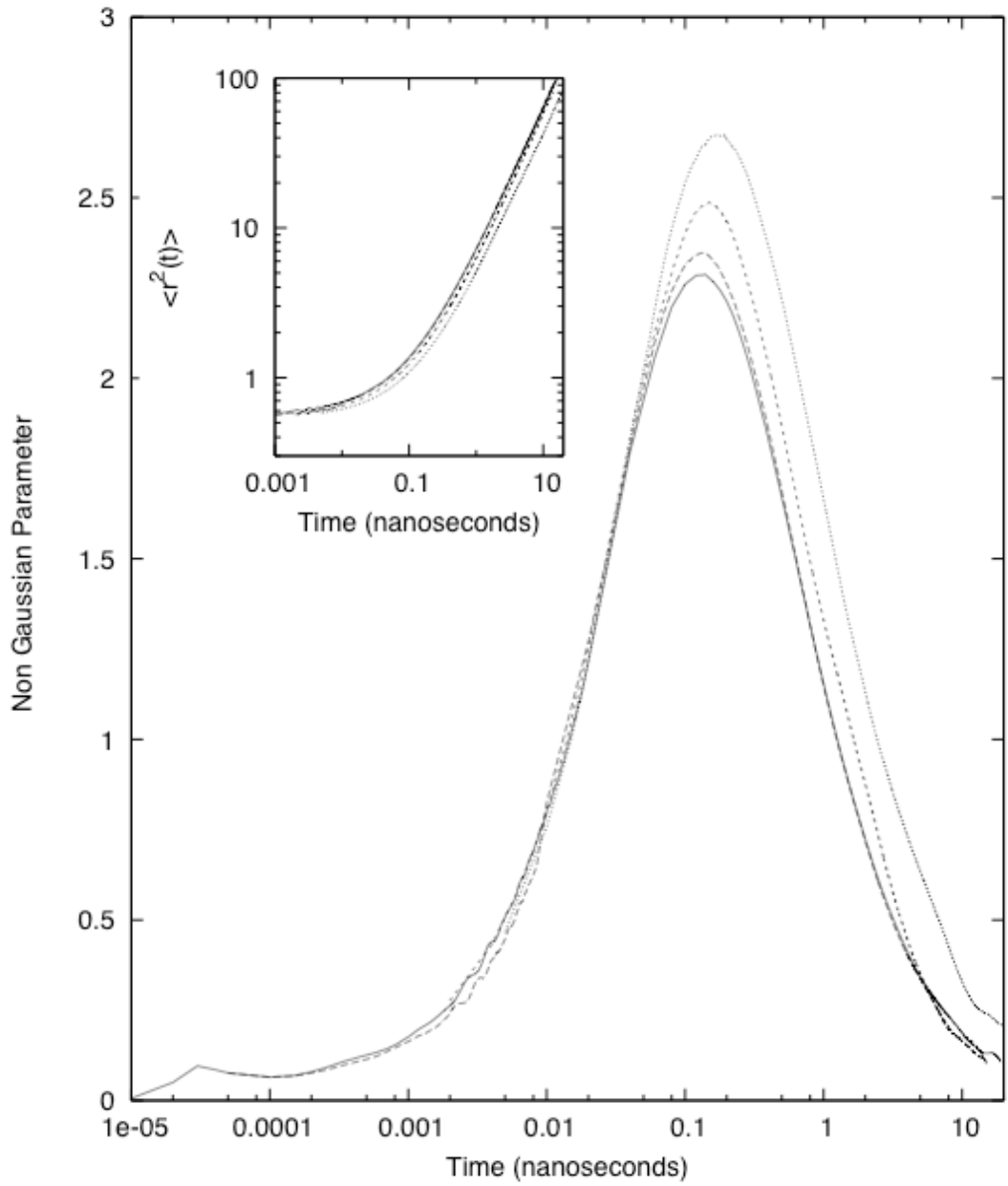

Figure 2a:

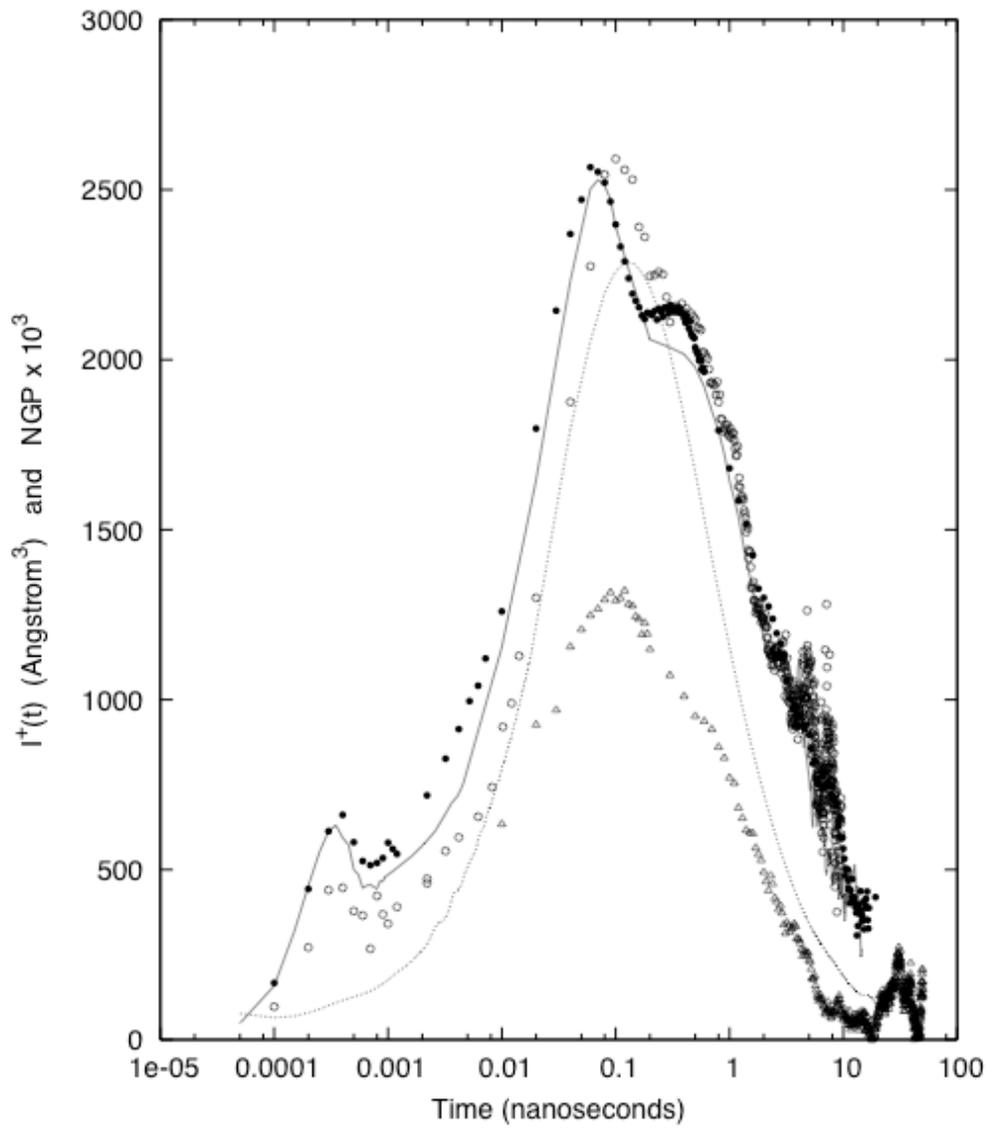

Figure 2b:

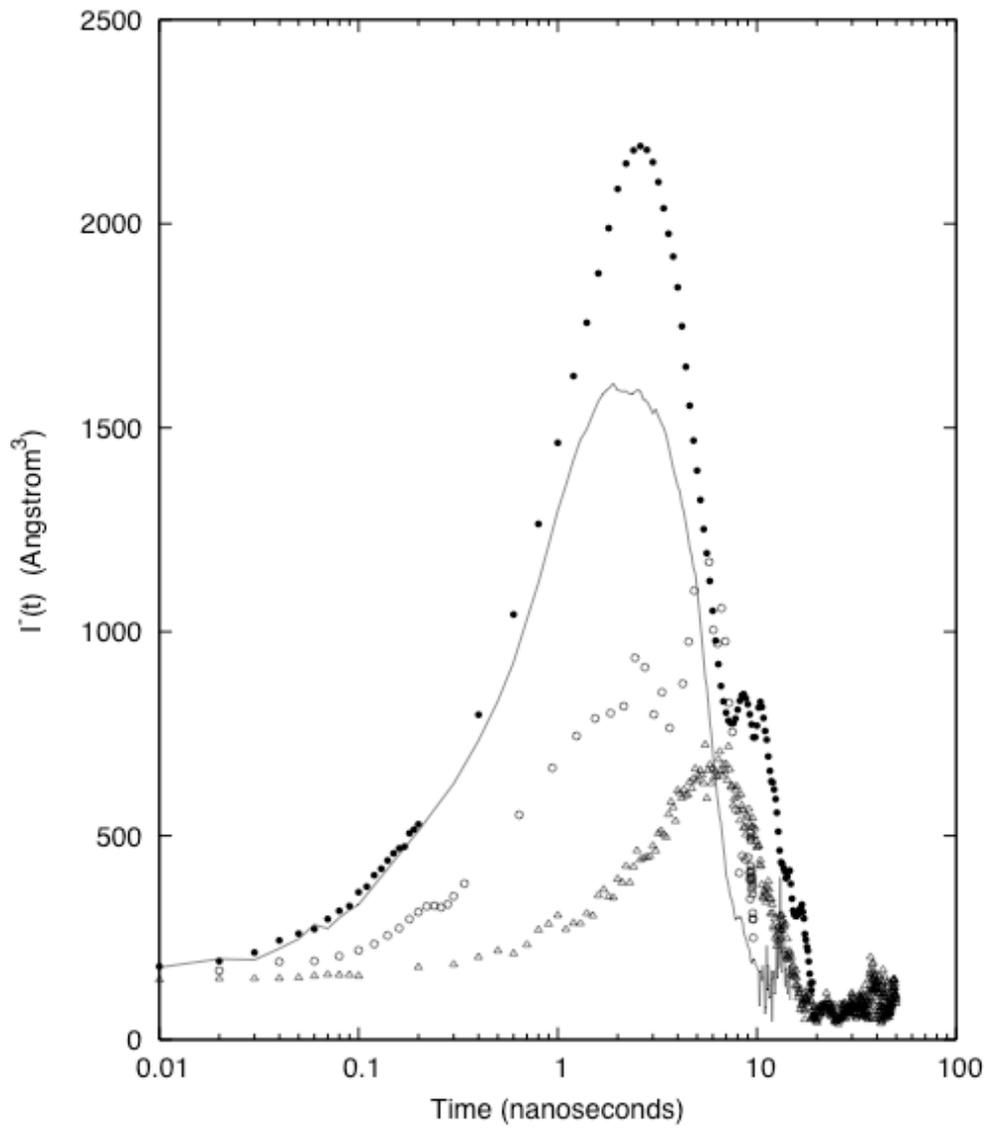

Figure 3:

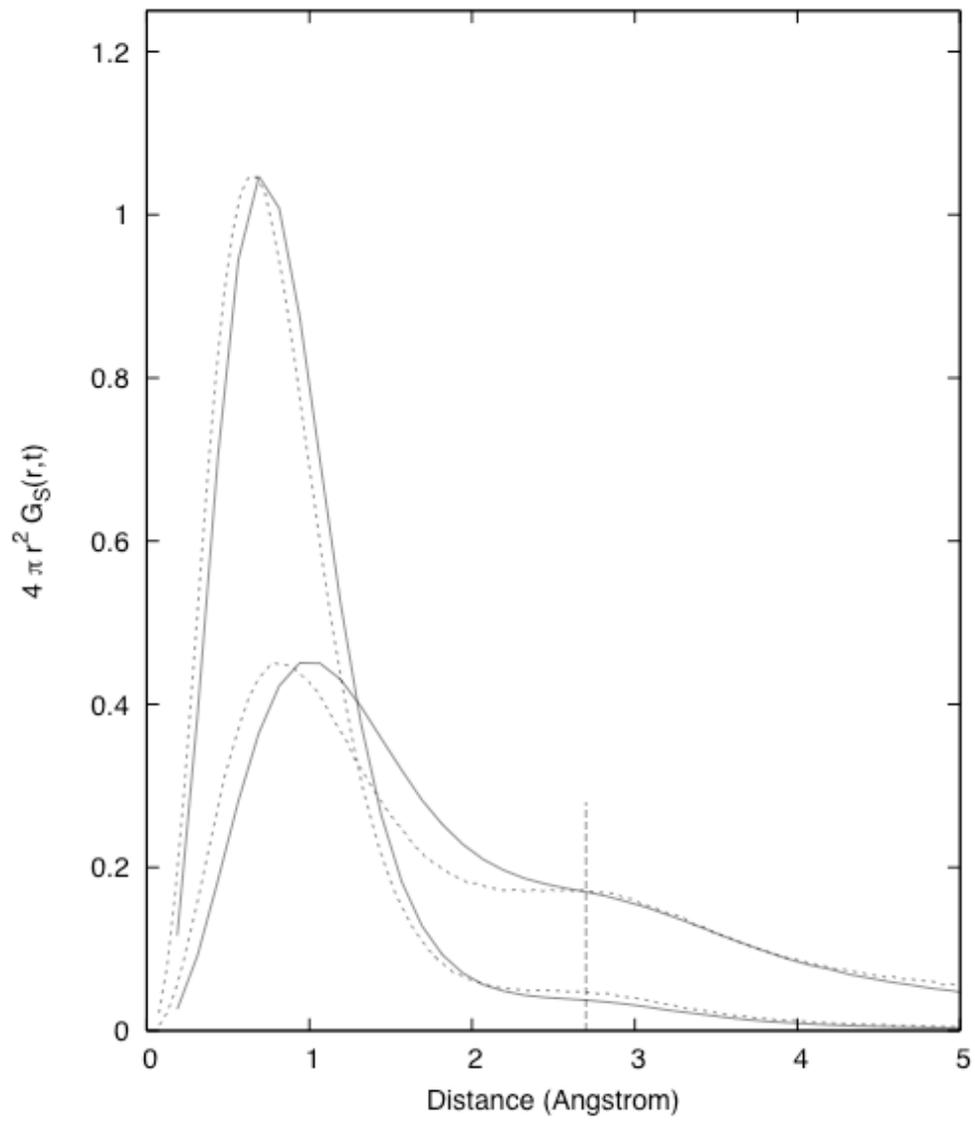